
\magnification=1200
\tolerance=10000
\font\titlea=cmb10 scaled\magstep1
\font\titleb=cmb10 scaled\magstep2
\rightline{UG-2/94}
\rightline{hepth@xxx/9404131}
\baselineskip=18pt
\vskip 1cm
\centerline{\titleb On N~=~1 Superconformal Algebra in}
\centerline{\titleb the Non-Critical Bosonic String Theory}
\vskip 2cm
\baselineskip=14pt
\centerline{Pablo M. Llatas\footnote{$^1$}{e-mail address:
Llatas@th.rug.nl}}
\medskip
\centerline{and}
\medskip
\centerline{Shibaji Roy\footnote{$^2$}{e-mail address: Roy@th.rug.nl}}
\bigskip
\centerline{\it Institute for Theoretical Physics}
\centerline{\it Nijenborgh 4, 9747 AG Groningen}
\centerline{\it The Netherlands}
\vskip 1cm
\baselineskip=18pt
\centerline{\titlea ABSTRACT}
\bigskip
In a recent work it has been shown that the bosonic strings
could be embedded into a special class of $N=1$ fermionic strings. We
argue that the superpartners of any physical state in the spectrum of
this fermionic string is non-physical. So, there is no supersymmetry in
the space of physical states and the embedding is, in this sense, ``trivial''.
We here propose two different constructions
as possible candidates of non-trivial embeddings of the non-critical bosonic
strings into some special class of $N=1$ fermionic strings of which one
is the non-critical NSR string. The BRST charge of the $N=1$ fermionic
strings in
both cases decompose as $Q_{N=1} = Q_B + {\tilde Q}$, where $Q_B$ is the BRST
charge of the bosonic string.
\vfill
\eject

Recently Berkovits and Vafa [1] have shown that the bosonic strings can be
regarded as a particular class of vacua for $N=1$ fermionic strings. Their
argument does not require a particular realization of bosonic strings but
depends on a general bosonic string vacuum i.e. a conformal field theory
with $c=26$. Although the amplitude computation in their case depends on a
special class of physical states known for the critical bosonic string,
the construction should also work for the non-critical bosonic strings. In
fact, a general cohomology argument has been given in ref.[2,3] which proves
an
isomorphism between the physical states of a general bosonic string and the
$N=1$ fermionic string. This particular $N=1$ fermionic string theory
possesses a
nonlinear $N=1$ supersymmetry and an action for this string
theory has been
constructed in ref.[4].

In this paper, we consider the non-critical ($c_M < 1$) bosonic
string theory and construct the $N=1$ generators in analogy with Berkovits
and Vafa. This construction gives an embedding of the non-critical bosonic
strings into a special class of $N=1$ fermionic strings. We then argue that
this
embedding is ``trivial'' in the sense that there is no supersymmetry in the
space of the physical states. We, however, propose two different
constructions of the $N=1$ generators from the fields of a non-critical bosonic
string which satisfy superconformal algebras with ${\hat c}={2\over 3}c=10$.
In the
first case, we use the hidden $N=2$ supersymmetry present in the non-critical
bosonic string theory [5,6,7] and `untwist' the energy-momentum tensor with
the
$U(1)$ current of the combined ghosts, matter and Liouville system. Note
that this was not possible for the critical bosonic string case in order
to keep the Lorentz invariance of the theory. The other $N=1$ generator, the
supercurrent, is constructed as a linear combination of the antighost field
$b$ and a modified BRST current. Unlike in the Berkovits-Vafa case, this
construction gives a possibility of a non-trivial embedding of non-critical
bosonic strings into a special class of $N=1$ fermionic strings. We point out
that this particular type of $N=1$ fermionic strings also possesses a nonlinear
$N=1$ supersymmetry. At this point it is natural to ask whether this special
class of $N=1$ fermionic strings has an interpretation in terms of the usual
NSR type
string having a linear $N=1$ supersymmetry. In other words, is it possible
to write down the $N=1$ generators as the energy-momentum tensor and the
supercurrent of a non-critical NSR string using some form of
bosonization? A few cases that we checked the answer to this question was
found to be negative. This indicates there is no clear target space
interpretation of this particular fermionic string.
These are the motivations for us to look for the second construction.
In this case, we
find two pairs of linear $N=1$ supermultiplets from the fields of a bosonic
string and can be regarded as the matter and Liouville supermultiplets of a
non-critical NSR string [8]. The $N=1$ generators $T_{N=1}$ and $G_{N=1}$ in
this case represent a novel realization of an $N=1$ superconformal algebra
(with ${\hat c} = {2\over 3}c = 10$) present in the non-critical bosonic
string theory. Moreover, we show that the energy-momentum tensor $T_{N=1}$
reduces
to the `untwisted' energy-momentum tensor of $c_M < 1$ bosonic string. This
result gives an indication that there is a possible embedding of non-critical
bosonic strings into non-critical NSR strings. We also show that the BRST
charge of the $N=1$ fermionic strings in both cases split up as $Q_{N=1} =
Q_B + {\tilde Q}$, $Q_B$ being the BRST charge of the bosonic string, which
is also quite suggestive of the type of embedding mentioned above.

We first describe briefly the Berkovits-Vafa [1] construction for the
non-critical bosonic strings. The total energy-momentum tensor of a
non-critical ($c_M < 1$) bosonic string is given as,
$$
{\tilde T}(z) = T_M(z) + T_L(z) + T^{gh}(z)\eqno(1)
$$
where
$$
\eqalign{T_M(z) &= -{1\over 2} :\partial\phi_M(z)\partial\phi_M(z):
+ i Q_M \partial^2\phi_M(z)\cr
T_L(z) &= -{1\over 2} :\partial\phi_L(z)\partial\phi_L(z): + i Q_L
\partial^2\phi_L(z)\cr
T^{gh}(z) &= -2 :b(z)\partial c(z): - :\partial b(z) c(z):\cr}\eqno(2)
$$
are the energy-momentum tensors of the matter, Liouville and the ghost
sectors with $\phi_M$, $\phi_L$, the matter and Liouville fields and
$(b,c)$,
the reparametrization ghost system having conformal weights 2 and $-1$
respectively. $2 Q_M$ and $2 Q_L$ are the background charges which satisfy
$Q_M^2 + Q_L^2 = -2$ since the total central charge of the combined
matter-Liouville system is 26. If we now `untwist' the energy-momentum
tensor (1)
by the ghost number current $J(z) = :c(z) b(z):$ as $T(z) = {\tilde T}(z)
-{1\over 2}\partial J(z)$, the conformal weight of the ghosts change from
($2,-1$) to (${3\over 2}, -{1\over 2}$) and can be regarded as the fermionic
matter of an $N=1$ fermionic string.

The $N=1$ generators are then given as,
$$
\eqalign{T_{N=1}(z) &= {\tilde T}(z) - {1\over 2}\partial J(z) + {1\over 2}
\partial^2(c_1\partial c_1)(z)\cr
G_{N=1}(z) &= b_1(z) + J_B(z) + {5\over 2} \partial^2 c_1(z)\cr}\eqno(3)
$$
where the `untwisted' ghosts with conformal weight $({3\over 2}, -{1\over 2})$
have been denoted as $(b_1, c_1)$ and $J_B(z) = :c_1(z)[T_M(z) + T_L(z) +
{1\over 2} T^{gh}(z)]:$ is the BRST current of the non-critical bosonic
string. The total derivative terms in (3) are required for the closure of
an $N=1$ superconformal algebra and the particular coefficients ensure that
the central charge is ${\hat c}=10$ so that (3) can be regarded as the matter
sector of an $N=1$ fermionic string theory. The physical content of this $N=1$
fermionic string theory can be obtained by looking at the cohomology classes
associated with the BRST charge
$$
\eqalign{Q_{N=1} &= \oint dz :\left[c(z) T_{N=1}(z) + \gamma(z) G_{N=1}(z)
+ b(z) c(z) \partial c(z) - b(z) \gamma^2(z)\right. \cr
&\left. \qquad\qquad\qquad + {3\over 2} c(z) \beta(z) \partial \gamma(z)
+ {1\over 2} c(z)
\partial \beta(z) \gamma(z)\right]:\cr}\eqno(4)
$$
where $(b, c)$ and $(\beta, \gamma)$ are respectively the diffeomorphism and
superdiffeomorphism ghost system of the $N=1$ string theory. It has been
shown in ref.[3], that there exists a homotopy operator $e^R$ where,
$$
\eqalign{ R &=- \oint dz :c_1(z)\left[\gamma(z) b(z) + {3\over 2} \partial c
(z) \beta(z) + c(z) \partial \beta(z)\right. \cr
&\left. \qquad\qquad\qquad\qquad +{1\over 2} \partial c_1(z) c(z) b(z)
+{1\over 4}\partial c_1(z)\beta(z)\gamma(z)\right]:\cr}\eqno(5)
$$
by which one can bring $Q_{N=1}$ to a simplified form through a similarity
transformation as,
$$
e^R~Q_{N=1}~e^{-R} = Q_{N=0} + Q_{top}\eqno(6)
$$
(Note that the numerical factors appearing in the operator $R$ in (5) are
different from those in ref.[3]).
Here $Q_{N=0} = \oint dz \left[c(z)\left(T_M(z) + T_L(z)\right) + :b(z) c(z)
\partial c(z):\right]$ is the BRST charge of the non-critical bosonic string
and $Q_{top} = \oint dz b_1(z) \gamma(z)$ is the BRST charge of a topological
sector. It is clear that $Q_{top}^2 = 0$ and $\lbrace Q_{N=0}, Q_{top}\rbrace
=0$. Also, since $Q_{top} c_1(z) = \gamma(z)$ and $Q_{top} \beta(z)
= -b_1(z)$, the only physical states in the cohomology of $Q_{top}$ are the
vacuum states. So, the physical states with respect to the cohomology of
$Q_{N=0} + Q_{top}$ have the form:
$$
|~{\rm bos}~>~\otimes~ |~0~>_{top}\eqno(7)
$$
where $|~{\rm bos}~>$ are the physical states of the bosonic string. From
(6) it is obvious that there is
a one-to-one map between the physical states with respect to the cohomology of
$Q_{N=0} + Q_{top}$ and those with respect to the cohomology of
$Q_{N=1}$. Since the operator $e^R~ \oint dz G_{N=1}(z)~ e^{-R}$ is fermionic
and it contains the fields in the topological sector $(b_1, c_1, \beta,
$ and $\gamma)$ we conclude that $\oint dz G_{N=1}(z)$ maps physical states
on the cohomology of $Q_{N=1}$ into the non-physical ones. It is,
therefore, clear that the supersymmetry is not realized in the space
of the physical states.

We here describe two new possiblities of embedding a non-critical bosonic
string into some special class of $N=1$ fermionic strings. In the first case
we use the hidden $N=2$ superconformal algebra in the non-critical bosonic
string theory. It is well-known [7] that the generator ${\tilde T}(z)$ as
given in (1) together with
$$
\eqalign{{\tilde G}^+(z) &= J_B(z) + a_1 \partial ( c \partial \phi_L)(z)
+ a_2 \partial (c \partial \phi_M)(z) + a_3 \partial^2 c(z)\cr
{\tilde G}^-(z) &= b(z)\cr
{\tilde J}(z) &= :c(z) b(z): - a_1 \partial \phi_L(z) - a_2 \partial
\phi_M(z)\cr}\eqno(8)
$$
where $J_B(z)$ is the BRST current given in (3) and $a_1$, $a_2$, $a_3$
are some arbitrary constants, satisfy a twisted $N=2$ superconformal algebra
with central charge $c^{N=2} = 6 a_3$ provided
$$
\eqalign{a_1 &= {1\over 4}\left[i Q_L (2 a_3 - 3) \pm Q_M (2 a_3 + 1)
\right]\cr
a_2 &= {1\over 4}\left[i Q_M (2 a_3 - 3) \mp Q_L (2 a_3 + 1)\right]\cr}
\eqno(9)
$$
This gives a one parameter family of twisted $N=2$ superconformal algebras
present in the non-critical bosonic string theory.

If we now `untwist' the energy-momentum tensor (1) with the $U(1)$ current
${\tilde J}(z)$ in (8) as,
$$
T(z) = {\tilde T}(z) - {1\over 2}\partial {\tilde J}(z)\eqno(10)
$$
then the conformal weight of ${\tilde G}^+(z)$ change from 1 to ${3 \over 2}$
and that of ${\tilde G}^-(z)$ change from  2 to ${3\over 2}$. Now, the new
generators,
$$
\eqalign{T_{N=1}(z) &\equiv T(z)\cr
G_{N=1}(z) &= a_4 {\tilde G}^-(z) + {\tilde G}^+(z)\cr}\eqno(11)
$$
where $a_4$ is another arbitrary constant, satisfy an $N=1$ superconformal
algebra with central charge ${\hat c} = {2\over 3} c = 10$ provided,
$$
\eqalign{a_1 &= {1\over 2}\left(i Q_L \pm 3 Q_M \right)\cr
         a_2 &= {1\over 2}\left(i Q_M \mp 3 Q_L \right)\cr
         a_3 &= {5\over 2}\cr
         a_4 &= 1\cr}\eqno(12)
$$
With these parameters, $T_{N=1}$, $G_{N=1}$ can be regarded as the matter
system of an $N=1$ fermionic string theory. We note that our energy-momentum
tensor $T_{N=1}$ does not require the total derivative term $\partial^2
(c\partial c)$ as in Berkovits-Vafa construction for the closure of the
algebra. Also, $G_{N=1}$ in our case matches exactly with Berkovits-Vafa
with $a_1 = a_2 = 0$. But $a_1 = a_2 = 0$ lead to inconsistencies in (12).
With the solution (12), the
energy-momentum tensor $T_{N=1}$ takes the form
$$
\eqalign{T_{N=1} &= -{1\over 2} :\partial \phi_M \partial \phi_M:
+ i {\hat Q}_M
\partial^2 \phi_M - {1\over 2} :\partial\phi_L \partial\phi_L: + i {\hat Q}_L
\partial^2 \phi_L\cr
&\qquad\qquad\qquad\qquad -{3\over 2} :b_1 \partial c_1: - {1\over 2}
:\partial b_1 c_1:\cr}\eqno(13)
$$
The new background charges $2{\hat Q}_M = {1\over 2}(5 Q_M \pm 3 i Q_L)$
and $2 {\hat Q}_L = {1\over 2}(5 Q_L \mp 3 i Q_M)$ also satisfy ${\hat Q}
^2_M + {\hat Q}^2_L = -2$. This, in fact, is the reason why the BRST charge
of the $N=1$ theory decompose as,
$$
Q_{N=1} = Q_B + {\tilde Q}\eqno(14a)
$$
where $Q_B = \oint dz :c\left[-{1\over 2} \partial\phi_M \partial\phi_M
+ i{\hat Q}_M \partial^2 \phi_M -{1\over 2} \partial\phi_L \partial\phi_L
+i{\hat Q}_L \partial^2 \phi_L - b\partial c\right]:$ is the BRST charge
of the bosonic string and
$$
\eqalign{{\tilde Q} &= \oint dz :\left[\gamma(z) G_{N=1}(z) -{3\over 2} c(z)
b_1(z)\partial c_1(z) - {1\over 2} c(z)\partial b_1(z) c_1(z) - b(z) \gamma^
2(z)\right. \cr
&\left. \qquad\qquad\qquad\qquad + {3\over 2}c(z)\beta(z)\partial \gamma(z)
+{1\over 2} c(z) \partial\beta(z)\gamma(z)\right]:\cr}\eqno(14b)
$$
We, however, do not find a homotopy operator by which (14a) can be brought
to a simplified form as in (6). This clearly indicates that we do not
have a decoupling of the fields $(b_1, c_1, \beta, \gamma)$ from the rest,
contrary to the Berkovits-Vafa
construction and (14) is quite suggestive that the non-critical bosonic string
could be embedded non-trivially into this $N=1$ fermionic string. Finally, we
note
that the matter content of this $N=1$ theory $(\partial \phi_L, \partial
\phi_M, b_1, c_1)$ have spins ($1, 1, {3\over 2}$, and $-{1\over 2}$) and so,
the $N=1$ supersymmetry is nonlinearly realized amongst them.

In the second case, we look for two pairs of linear $N=1$ supermultiplets
($\partial \phi_1, \psi_1$) and ($\partial\phi_2, \psi_2$) constructed
from the matter,
Liouville and the reparametrization ghosts of a non-critical bosonic string
and consider them as the matter and the Liouville supermultiplets of a
non-critical NSR string. The energy-momentum tensor and the supercurrent for
a non-critical NSR string are given as
$$
\eqalignno{T_{N=1}(z) &= -{1\over 2} :\partial\phi_1(z) \partial\phi_1(z):
+i Q_1 \partial^2 \phi_1(z) -{1\over 2} :\partial\phi_2(z) \partial
\phi_2(z): + i Q_2 \partial^2\phi_2(z)\cr
&\qquad\qquad\qquad\qquad -{1\over 2} :\psi_1(z)\partial\psi_1(z): -
{1\over 2} :\psi_2(z)\partial\psi_2(z): &(15a)\cr
G_{N=1}(z) &= i\partial \phi_1(z)\psi_1(z) + 2 Q_1 \partial \psi_1(z)
+i\partial\phi_2(z)\psi_2(z) + 2 Q_2 \partial\psi_2(z) &(15b)\cr}
$$
where $2 Q_1$ and $2 Q_2$ are the background charges associated with the
matter and the Liouville sector of a non-critical NSR string.

These generators satisfy an $N=1$ superconformal algebra with central charge
${\hat c}= {2\over 3}c=10$, provided the background charges satisfy
$$
Q_1^2 + Q_2^2 = -1 \eqno(16)
$$
The basic OPEs are defined as,
$$
\eqalignno{<\partial\phi_1(z) \partial\phi_1(w)> &= <\partial\phi_2(z)
\partial\phi_2(w)> = -{1\over {(z-w)^2}} &(17a)\cr
<\psi_1(z) \psi_1(w)> &= <\psi_2(z) \psi_2(w)> = {1\over {(z-w)}} &(17b)\cr}
$$
Let us now consider an energy-momentum tensor of the form,
$$
\eqalign{T(z) &= -{1\over 2} :\partial\phi_M(z) \partial\phi_M(z): + i Q_M
\partial^2 \phi_M(z) - {1\over 2} :\partial\phi_L(z) \partial\phi_L(z):
+ i Q_L \partial^2\phi_L(z)\cr
& \qquad\qquad\qquad -(k+{1\over 2}) :b_1(z) \partial c_1(z): - (k-{1\over 2})
:\partial b_1(z) c_1(z):\cr}\eqno(18)
$$
where $\partial \phi_M(z)$, $\partial\phi_L(z)$ are two bosonic conformal
fields with weight 1 and $b_1(z)$, $c_1(z)$ are two fermionic conformal fields
with weight $(k+{1\over 2})$ and $(-k+{1\over 2})$ respectively. We introduce
two more conformal fields $:e^{\pm i\lambda(\phi_M - i \phi_L)}:$ where
$\lambda\ge 0$, with conformal weights $\mp \lambda(Q_M-iQ_L)$ with respect to
(18). The basic OPEs in this case are
$$
\eqalign{<\partial\phi_L(z) \partial\phi_L(w)> &= <\partial\phi_M(z)
\partial\phi_M(w)> = -{1\over {(z-w)^2}}\cr
<b_1(z) c_1(w)> &= {1\over{(z-w)}}\cr}\eqno(19)
$$
We here note that the two fermionic fields $:b_1 e^{i\lambda(\phi_M -
i\phi_L)}:$ and $:c_1 e^{-i\lambda(\phi_M -i\phi_L)}:$ have the same conformal
weights equal to ${1\over 2}$ with respect to (18) provided $k = \lambda
(Q_M-iQ_L)$. We, therefore, define two fermionic fields,
$$
\eqalign{\psi_1(z) &= \left[:b_1(z) e^{i\lambda(\phi_M - i\phi_L)}(z):
+{1\over 2} : c_1(z) e^{-i\lambda(\phi_M -i\phi_L)}(z):\right]\cr
\psi_2(z) &= i \left[:b_1(z) e^{i\lambda(\phi_M -i\phi_L)}(z): -
{1\over 2} :c_1(z) e^{-i\lambda(\phi_M - i\phi_L)}(z):\right]\cr}\eqno(20)
$$
consistent with the basic OPEs (17).

Now we define two bosonic fields with conformal weight 1 with respect to
(18) as,
$$
\eqalign{\partial\phi_1(z) &= t_1 \partial \phi_M(z) + t_2 \partial \phi_L(z)
+ \lambda(i t_1 + t_2) :b_1(z) c_1(z):\cr
\partial\phi_2(z) &= g_1 \partial\phi_M(z) + g_2 \partial\phi_L(z) +\lambda
(i g_1+ g_2):b_1(z) c_1(z):\cr}\eqno(21)
$$
where $t_1, t_2, g_1, g_2$ are arbitrary parameters. (20) and (21) give
consistent basic OPEs among themselves if the parameters satisfy,
$$
\eqalign{(t_1 + i t_2) (t_1 - i t_2) +\lambda^2 (t_1 - i t_2)^2 &= 1\cr
(g_1 + i g_2) (g_1 - i g_2) + \lambda^2 (g_1 - i g_2)^2 &= 1 \cr
g_1 t_1 + g_2 t_2 + \lambda^2 (g_1 t_1 - g_2 t_2 - i g_1 t_2 - i g_2 t_1)
&=0\cr}\eqno(22)
$$
Substituting (20) and (21) in (15a) we find that $T_{N=1}$ matches precisely
with (18) provided the following relations are satisfied,
$$
\eqalignno{g_1^2 + t_1^2 + \lambda^2 &= 1 &(23a)\cr
g_2^2 + t_2^2 -\lambda^2 &= 1 &(23b)\cr
g_1 g_2 + t_1 t_2 - i\lambda^2 &= 0 &(23c)\cr
g_2 Q_2 + t_2 Q_1 &= Q_L &(23d)\cr
g_1 Q_2 + t_1 Q_1 &= Q_M &(23e)\cr}
$$
alongwith $k = \lambda(Q_M - i Q_L)$. We note that (22) are not new
conditions but follow from (23). We here give the solutions of
(23a)--(23e) for general $k$ and $\lambda$,
$$
\eqalign{g_1 &= \pm {i\over 2} Q_1\left({\lambda\over k} + {k\over \lambda}
-\lambda k\right) + {1\over 2} Q_2 \left({\lambda \over k} - {k\over \lambda}
+\lambda k\right)\cr
g_2 &= \mp {1\over 2} Q_1 \left( -{\lambda \over k} + {k \over \lambda}
+\lambda k\right) - {i\over 2} Q_2 \left({\lambda \over k} + {k\over \lambda}
+\lambda k\right)\cr
t_1 &= {1\over 2} Q_1 \left({\lambda \over k} - {k\over \lambda} + \lambda k
\right) \mp {i\over 2} Q_2 \left({\lambda \over k} + {k\over \lambda} -
\lambda k\right)\cr
t_2 &= -{i \over 2} Q_1 \left( {\lambda \over k} + {k\over \lambda} +
\lambda k\right) \pm {1\over 2} Q_2 \left(-{\lambda\over k} +{k\over \lambda}
+\lambda k\right)\cr}\eqno(24)
$$
We here point out that the two solutions in (24) are not independent as they
can be obtained by $g_1 \leftrightarrow t_1$ ; $g_2 \leftrightarrow t_2$;
$Q_1 \leftrightarrow Q_2$; $\pm \leftrightarrow \mp$ which is a symmetry of
(23) and interchange the role of $\partial\phi_1$ and $\partial\phi_2$ in (21).
Also, using (24) we get from (23d) and (23e)
$$
\eqalignno{Q_L &= {i\over 2}({\lambda \over k} + {k\over \lambda} +
\lambda k) &(25a)\cr
Q_M &= -{1\over 2}({\lambda \over k} - {k\over \lambda} +\lambda k) &(25b)\cr
Q_L^2 + Q_M^2 &= -1-k^2 &(25c)\cr}
$$
So, for general $k = \lambda(Q_M - iQ_L)$ we find an exact correspondence
between the two energy-momentum tensors (15a) and (18). Note, however, that
for $k=1$ (25c) is precisely the background charge condition for a
non-critical bosonic string and (18) can be recognized as the `untwisted'
(untwisted by the ghost number current $J(z) = :c(z) b(z):$) energy-momentum
tensor. In this case $Q_L = i (\lambda + {1\over {2\lambda}})$ and $Q_M =
({1\over {2\lambda}} - \lambda)$. For $\lambda = \sqrt{{q\over {2p}}}$, they
describe the background charges of ($p,q$) minimal model coupled to gravity.
This, therefore, proves an equivalence between the `untwisted'
energy-momentum tensor of a non-critical bosonic string with that of the
non-critical NSR string.

The other $N=1$ generator, the supercurrent in (15b) expressed in terms of
fields in the bosonic string is given as,
$$
\eqalign{G_{N=1} &= (Q_1 + i Q_2):\left[(3\lambda + {1\over \lambda})
\partial \phi_L b_1 + i (3\lambda -{1\over \lambda})\partial\phi_M b_1
+ 4 \partial b_1\right] e^{i\lambda(\phi_M - i \phi_L)}:\cr
&\qquad +(Q_1 - i Q_2) :\left[ -{\lambda \over 2} \partial \phi_L c_1
-{i\over 2}\lambda \partial \phi_M c_1 + \partial c_1\right] e^{-i\lambda
(\phi_M - i\phi_L)}:\cr}\eqno(26)
$$
It is straightforward to check that the generators (18) with $k=1$ and (26)
indeed satisfy an $N=1$ superconformal algebra with ${\hat c} = {2\over 3}c
=10$. We would like to point out that if we consider ($\partial\phi_1,
\psi_2$) and ($\partial\phi_2, \psi_1$) as the supermultiplets then the
generator $G_{N=1}$ takes a slightly different form:
$$
\eqalign{G_{N=1} &= (Q_1 - i Q_2):\left[ 3 i \lambda \partial\phi_L b_1
-3 \lambda \partial\phi_M b_1 + 2 i \partial b_1 \right] e^{i \lambda(\phi
_M- i\phi_L)}:\cr
&\qquad +{1\over 2}(Q_1 + i Q_2):\left[ i (\lambda - {1\over \lambda})
\partial
\phi_L c_1 - (\lambda + {1\over \lambda})\partial \phi_M c_1\right] e^{
-i \lambda (\phi_M - i \phi_L)}:\cr}\eqno(27)
$$
Eqs.(18), (26) or (27) therefore gives a novel realization of an $N=1$
superconformal algebra in the non-critical bosonic string theory.

The BRST charge in this case can be written as in (4) with $T_{N=1}$ as
given in (18) with $k=1$ and $G_{N=1}$ as given either in (26) or in (27).
Since (18) is the `untwisted' energy-momentum tensor of a non-critical
bosonic string, it is clear that the BRST charge would split up again as,
$$
Q_{N=1} = Q_B + {\tilde Q}\eqno(28)
$$
where ${\tilde Q}$ is again as given by (14b) with $G_{N=1}$ given in (26)
or (27). We would like to point out here that this splitting is not obvious
because ${\tilde Q}$ does not contain any term present in $Q_B$.
However, in this case also we do not find a homotopy operator to
bring $Q_{N=1}$ into a simplified form as in (6). This is quite expected
since the physical states of an NSR string are not of the type (7).

To conclude, we have argued that the Berkovits-Vafa embedding of bosonic
strings into special class of $N=1$ fermionic strings is ``trivial''
as there is no supersymmetry in the space of the physical states of
this $N=1$ fermionic string theory. We have
presented two separate constructions as possible candidates for non-trivial
embeddings of non-critical bosonic strings into $N=1$ fermionic strings.
In one
of the cases the $N=1$ fermionic string is the usual NSR string. In both
cases that
we have considered the BRST charge splits up as $Q_{N=1} = Q_B + {\tilde Q}$,
where $Q_B$ is the BRST charge of the non-critical bosonic string. We,
however, do not find any homotopy operator by which $Q_{N=1}$ can be brought
into $Q_B + Q_{top}$ as happened in Berkovits-Vafa case. This clearly shows
that we do not have a trivial decoupling of ($b_1, c_1, \beta, \gamma$) from
the bosonic string and we may have a non-trivial embedding. In fact, the
splitting of the BRST charges as given above suggests that there could be
a similarity transformation by which $Q_{N=1}$ would decompose into two
nilpotent and mutually anticommuting BRST charges, where one of them
obviously is the bosonic BRST charge. This could simplify the cohomology
analysis of the $N=1$ fermionic string theory. These
questions are presently under investigation.
\vskip 1cm
\noindent{\titlea Acknowledgements:}
\medskip
We would like to thank E. Bergshoeff, N. Berkovits,
J. M. Figueroa-O'Farrill, H. Ishikawa, M. Kato, J. M. F. Labastida, A. V.
Ramallo, M. de Roo, J. Sanchez-de-Santos and especially H. J. Boonstra for
discussions at various stages of the
work. We also thank K. Thielemans for providing us his latest version of the
Mathematica package OPEconf and ref.[9] which was used in some OPE
computations. The
work of S. R. was performed as part of the research program of the ``Stichting
voor Fundamenteel Onderzoek der Materie'' (FOM). The work of P. M. Ll. is
supported by the ``Human Capital and Mobility Program'' of the European
Community.
\vfill
\eject
\noindent{\titlea References:}
\item{1.} N. Berkovits and C. Vafa, {\it On the Uniqueness of String
Theory}, preprint HUTP-93/A031, KCL-TH-93-13, hep-th/9310170 (1993).
\item{2.} J. M. Figueroa O'Farrill, Phys. Lett. B321 (1994) 344.
\item{3.} H. Ishikawa and M. Kato, {\it Note on N=0 String as N=1 String},
preprint UT-Komaba/93-23, hep-th/9311139 (1993).
\item{4.} F. Bastianelli, Phys. Lett. B322 (1994) 340.
\item{5.} B. Gato-Rivera and A. Semikhatov, Phys. Lett. B288 (1992) 295.
\item{6.} M. Bershadsky, W. Lerche, D. Nemeschansky and N. P. Warner,
Nucl. Phys. B401 (1993) 304.
\item{7.} S. Panda and S. Roy, Phys. Lett. B317 (1993) 533.
\item{8.} M. Kato and S. Matsuda, Adv. Stud. Pure Math. 16 (1988) 205.
\item{9.} K. Thielemans, Int. Jour. Mod. Phys. C2 (1991) 787.

\end